\documentstyle[12pt,epsfig]{article}
\def\NPB{{\em Nucl. Phys.} B}
\def\PLB{{\em Phys. Lett.}  B}

\def\PRD{{\em Phys. Rev.} D}
\def\ZPC{{\em Z. Phys.} C}
\begin{document}
\title{Multi-parameter fits to the $t\bar{t}$ threshold observables at a future $e^+e^-$ linear collider}
\author{Manel Martinez$^a$, Ramon Miquel$^b$ \\
{\em $^a$ Institut de F\'\i sica d'Altes Energies, Univ.~Aut.~de Barcelona}\\ 
{\em E-08390 Bellaterra (Barcelona) Spain}\\
{\em $^b$ Lawrence Berkeley National Laboratory, Physics Division} \\ 
{\em 1 Cyclotron Road, Berkeley, CA 94720, USA} }
\maketitle
\baselineskip=11.6pt
\begin{abstract}
A realistic study of the physics reach of a $t\bar{t}$ threshold scan at a future $e^+e^-$ linear collider is presented.
The results obtained take into account experimental and, to a large extent,
theoretical systematic errors, as well as beam effects. Because of
the large correlations between the physical parameters that can be extracted from the threshold scan, a multi-parameter
fit is seen as mandatory. It is shown that the top mass, the top width and $\alpha_s(M_Z)$ can
be extracted simultaneously with uncertainties around 20~MeV, 30~MeV and 0.0012, respectively, while
the top Yukawa coupling can be measured, with the previous three parameters, to an uncertainty of about 35\%, after
assuming an external prior on $\alpha_s$ of $\pm 0.001$.
\end{abstract}
\baselineskip=14pt
\section{Introduction}
The study of the top threshold scan observables in a $e^+ e^-$ linear collider environment was pointed 
out long time ago~\cite{history} as a potentially high precision strategy for the determination of the top mass and 
eventually other relevant parameters such as the strong coupling constant, the top quark width and 
the top Yukawa coupling. Since then many studies
with increasing levels of complexity have been performed 
to obtain a quantitative estimate of the attainable accuracy.

Earlier linear collider studies of the top threshold focused on the determination of the top
quark mass~\cite{Bagliesi, Igo, Comas, japan, usa}. A strong correlation between the top mass
and the strong coupling constant, $\alpha_s(M_Z)$, was noticed, so that both quantities had to
be measured at the same time, through a simultaneous two-parameter 
fit~\cite{Bagliesi, Igo}. The correlation limited the experimental
precision that could be achieved for the top mass to about 300~MeV. On top of it, when
next-to-next-to-leading corrections to the $\bar{t}t$ cross section at threshold where 
computed~\cite{teubner-98}, they were found to be large and to perturb the determination of the
mass at the level of about 500~MeV.

In 1999 there was a
substantial breakthrough when two new definitions of top mass
(``potential subtracted''~\cite{PS-mass} and 
``1S'' ~\cite{1S-mass} ) were proved to be much less
sensitive to higher order corrections than the pole mass used previously. As a welcome side
effect, correlations between $\alpha_s$ and these new masses were found to be much 
reduced~\cite{sitges-99}, so that a determination of $m_t$ with less than 100~MeV
experimental error and about $100-150$~MeV theoretical uncertainty became 
feasible~\cite{sitges-99}. 
The decrease in the correlation can be understood from the fact that the most sensitive information comes from the 
position of the $1S$ resonance peak (although severely smeared after initial state radiation (ISR) 
and beam effects), which is at an energy $E_{1S} = 2 m_t - V_{t\bar{t}}(\alpha_s)$,
where $m_t$ stands for the top pole mass, used previously, and $V_{t\bar{t}}(\alpha_s)$ stands for the 
$t\bar{t}$ binding potential, and depends almost linearly on $\alpha_s$, therefore heavily correlating both parameters.

In order to help disentangle these two variables, the top momentum distribution due to the top 
Fermi motion was advocated already several years ago \cite{momentum-kuhn,chapas}. 
Simulation studies showed that the peak position 
of that distribution was quite robust against ISR and beam effects and changed linearly with the top 
mass while being insensitive to the strong coupling constant, providing, therefore, an additional handle for 
the disentangling of both variables \cite{chapas,Igo}.

Finally, in addition, the use of the top forward-backward charge asymmetry was suggested in the past 
as a way of getting direct information on the top quark width~\cite{afb-sumino,afb-kuhn}. 
The large width of the top quark is 
responsible for the overlap between the $1S$ and $1P$ states whose interference causes a forward-backward charge
asymmetry. Simulation studies showed that the measurement was feasible, although the attainable 
accuracy was rather limited \cite{Comas}.

In all these observables some modest sensitivity to the influence of the top Yukawa coupling, entering the $t\bar{t}$ 
potential was expected. Again, Monte Carlo studies showed that indeed the sensitivity was quite low \cite{Bagliesi}.

For the top width and the top Yukawa coupling studies, the approach followed so far was a single parameter 
determination assuming no relevant uncertainty in the other parameters (top mass and strong coupling constant). 

The study presented in this paper continues the work of ref.~\cite{sitges-99}, extending it and completing it.
It completes it because it includes not just the cross section but also the other two relevant observables,
momentum distribution and forward-backward asymmetry. It extends it because it explores the feasibility of 
extracting information about all the relevant input parameters simultaneously, 
that is, not just the top mass and the strong 
coupling constant but also the top width and the top Yukawa coupling.

The outline of the paper is as follows: in section 2 the conditions in which 
the present analysis has been developed are summarized. In section 3 the 
results for a ``standard'' two parameter fit to the top mass and the strong 
coupling constant are presented and the sources of the correlations 
obtained are discussed. The case for a multi-parameter approach is then 
presented in section 4 and the technical solution is described. 
Using this multi-parameter approach, section 5 deals with the discussion on 
the actual sensitivity to the top quark width and section 6 with the prospects
for a measurement of the top Yukawa coupling.
Finally, section 7 summarizes the conclusions of this study.
\section{Simulation Input}
\label{experiment}
For the present study, the simulation conditions have been assumed to be identical to the ones used 
in~\cite{sitges-99}, namely, the TESLA beam conditions of ref.~\cite{TESLA} were assumed and the detector effects
were simulated using the {\tt SIMDET} algorithm as described in ref.~\cite{SIMDET}.

The $t \bar{t}$ observables (cross section, top momentum distribution and forward-backward charge asymmetry) have 
been computed using the {\tt TOPPIK} code~\cite{momentum-kuhn, afb-kuhn, TOPPIK, kuhn-yuk} including 
the latest theoretical predictions as discussed 
in ref.~\cite{teubner2001}~\footnote{A very recent update~\cite{hoang2002} of ref.~\cite{teubner2001} results in
shifts in the cross section of up to 1.5\%, well within the assumed theoretical error. The new spin-independent
$1/m^2$ potentials
in~\cite{hoang2002} can be seen to agree with those obtained by Pineda in~\cite{pineda}, while those
in~\cite{teubner2001} disagreed.}.
For the studies presented in this work, the $1S$ mass definition has been used.
The actual values of the input parameters used in the calculation are $m_t(1S)=175$ GeV, $\alpha_s(M_Z)=0.120$, 
$M_H=120$ GeV and the top width and Yukawa coupling as predicted in the Minimal Standard Model.
For the experimental selection studies, the signal and the relevant backgrounds have been generated using 
{\tt PYTHIA}~\cite{PYTHIA}. The event simulation is discussed in detail in ref.~\cite{Aurelio}. 

For the cross section analysis, purely hadronic decays of the $t\bar{t}$ system, together with events in which 
only one of the top quarks decays 
hadronically are used. This results in an event selection efficiency of 41.2 \% 
(over the complete $t\bar{t}$ sample) with an estimated systematic uncertainty in the selection efficiency of
3\% and a remaining
background 
cross section of about 0.0085 pb. The same sample is used for the study of the top momentum distribution and, hence, 
the efficiency remains the same, with an estimated systematic uncertainty in the peak value of the momentum 
distribution of about 4\%.
For the forward backward asymmetry measurement, only the sample in which one of the top decays semileptonically to an
electron or a muon, allowing to tag easily the top charge, is used. This results in an event selection efficiency of about 
14~\% with negligible background and systematic uncertainties.

For the first time, theoretical uncertainties have been included in the fit. Following ref.~\cite{teubner2001} a 3\%
uncertainty in the total cross section, common to all center-of-mass energy points, has been assumed. 
This has to be considered as only a
first attempt at quantifying the influence of the theoretical error in the results~\footnote{For instance, we have 
noticed that assuming, instead, that the 3\% theoretical error is totally uncorrelated between energy points (which
seems rather unlikely), the effect of the theoretical error becomes much more prominent.}.
No estimate of theoretical systematics is available for either the top momentum
distribution or the forward-backward charge asymmetry. However, as it will be shown below, these two observable have a
rather limited weight in the final results.

The threshold scan has been assumed to consist of a luminosity of 300~fb$^{-1}$ uniformly distributed in 10 scan 
points: one of them well below threshold for a direct background determination and the other 9 distributed 
symmetrically around the $m_t(1S)$ value with a 1 GeV spacing between each 
other~\footnote{Some studies carried out in the past using just the cross section and fitting only the top mass and 
the strong coupling constant showed already that some modifications of the scanning strategy could allow to decrease
some parameter errors~\cite{sitges-99}. 
However, when dealing with four parameter fits, as we are doing here, the optimization of the scan strategy is not so
obvious and for the moment it has not been attempted.}.
Since the top mass will only be known with a moderate precision
at the time the top threshold scan starts at a linear collider, it will be
impossible to choose the center-of-mass energy points to lie precisely at the values we have chosen relative to $2 m_t$.
However, it has been checked that assuming a prior precision on $m_t$ of around 500~MeV (coming from measurements at
LHC or in the continuum region at a linear collider), the effect of not choosing the optimal values for $\sqrt s$ is
very small.

\begin{figure}
\epsfxsize=15cm
\leavevmode
\centering
\epsffile{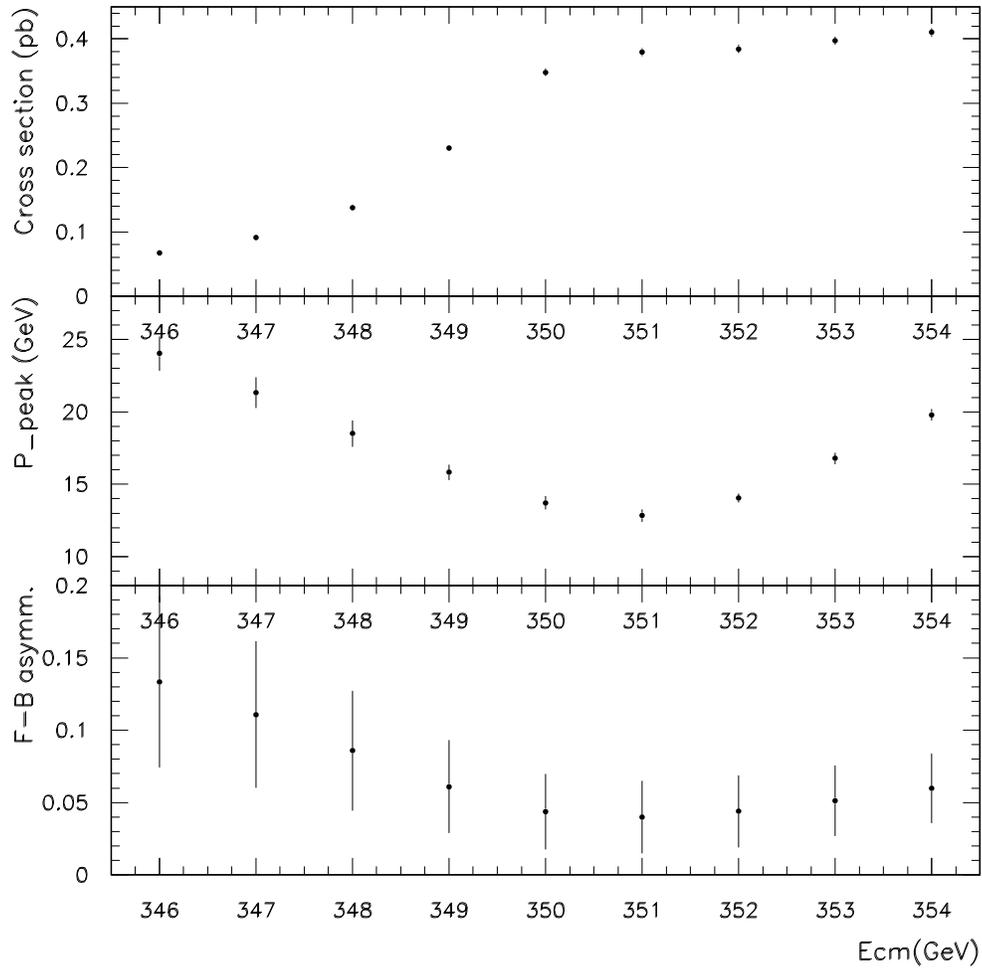}
\caption{The expected scan results for the cross section, the peak of 
the top momentum distribution and the forward-backward charge asymmetry in 
the conditions described in the text.}
\label{fig:Experiment}
\end{figure}

The expectations
for the three observables studied here are shown in fig.~\ref{fig:Experiment} together with the corresponding 
expected experimental errors. It is clear from that figure that, while the measurement of cross section will be very
precise, the peak of the momentum distribution will be determined with a moderate precision and the 
forward-backward asymmetry with a rather low precision. These uncertainties should be kept in mind when analyzing 
the sensitivity of each observable to the input parameters in the next sections. Both things put together will provide 
a feeling for which measurements are actually important for the determination of each one of the parameters discussed.
%
\section{The Top Mass and the Strong Coupling Constant}
To start with, a two parameter fit, with $m_t$ and $\alpha_s(M_Z)$, is
performed, as in ref.~\cite{sitges-99} but now using the larger integrated luminosity mentioned
in the previous section and including also the information in the forward-backward
charge asymmetry ($A_{FB}$)
and position of the peak of the top momentum distribution, as outlined also
in section~\ref{experiment}. 

\begin{figure}
\epsfxsize=15cm
\leavevmode
\centering
\epsffile{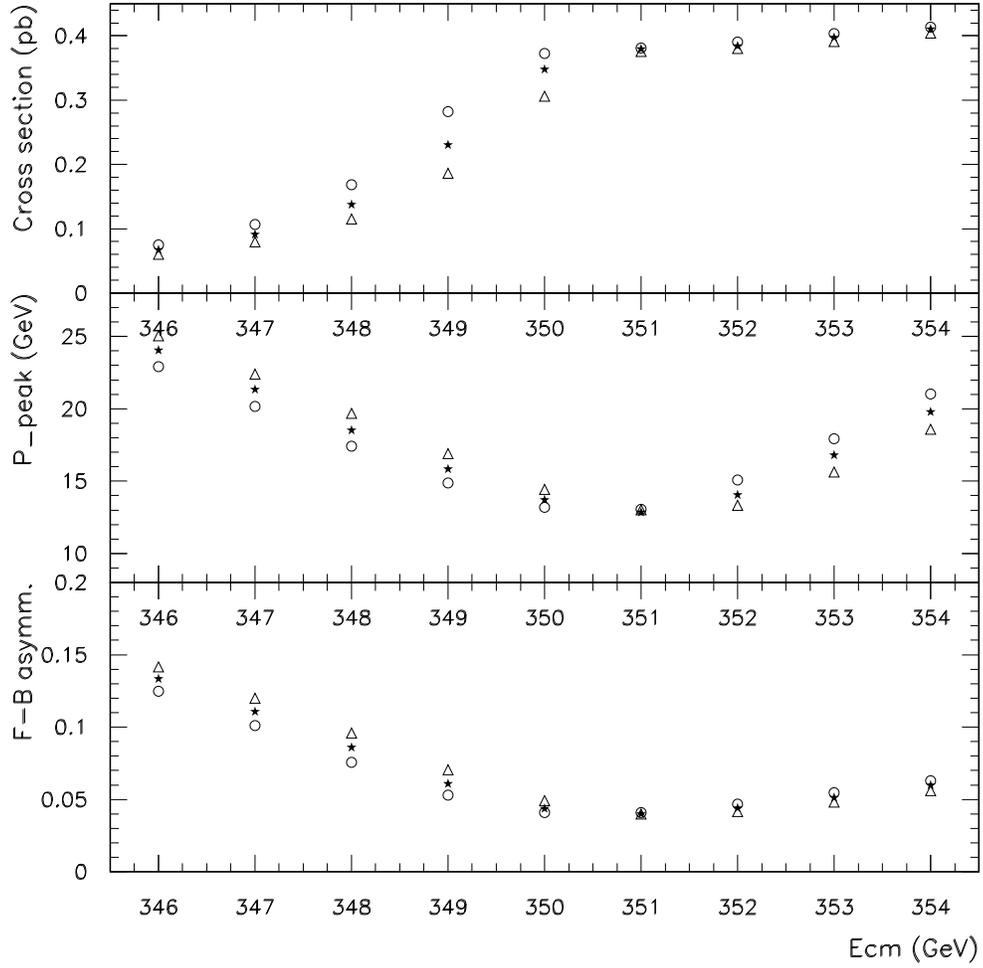}
\caption{Sensitivity of the observables to the top mass. The different markers 
correspond to $\Delta m_t = 200$ MeV intervals.}
\label{fig:sensitivity-mt}
\end{figure}

\begin{figure}
\epsfxsize=15cm
\leavevmode
\centering
\epsffile{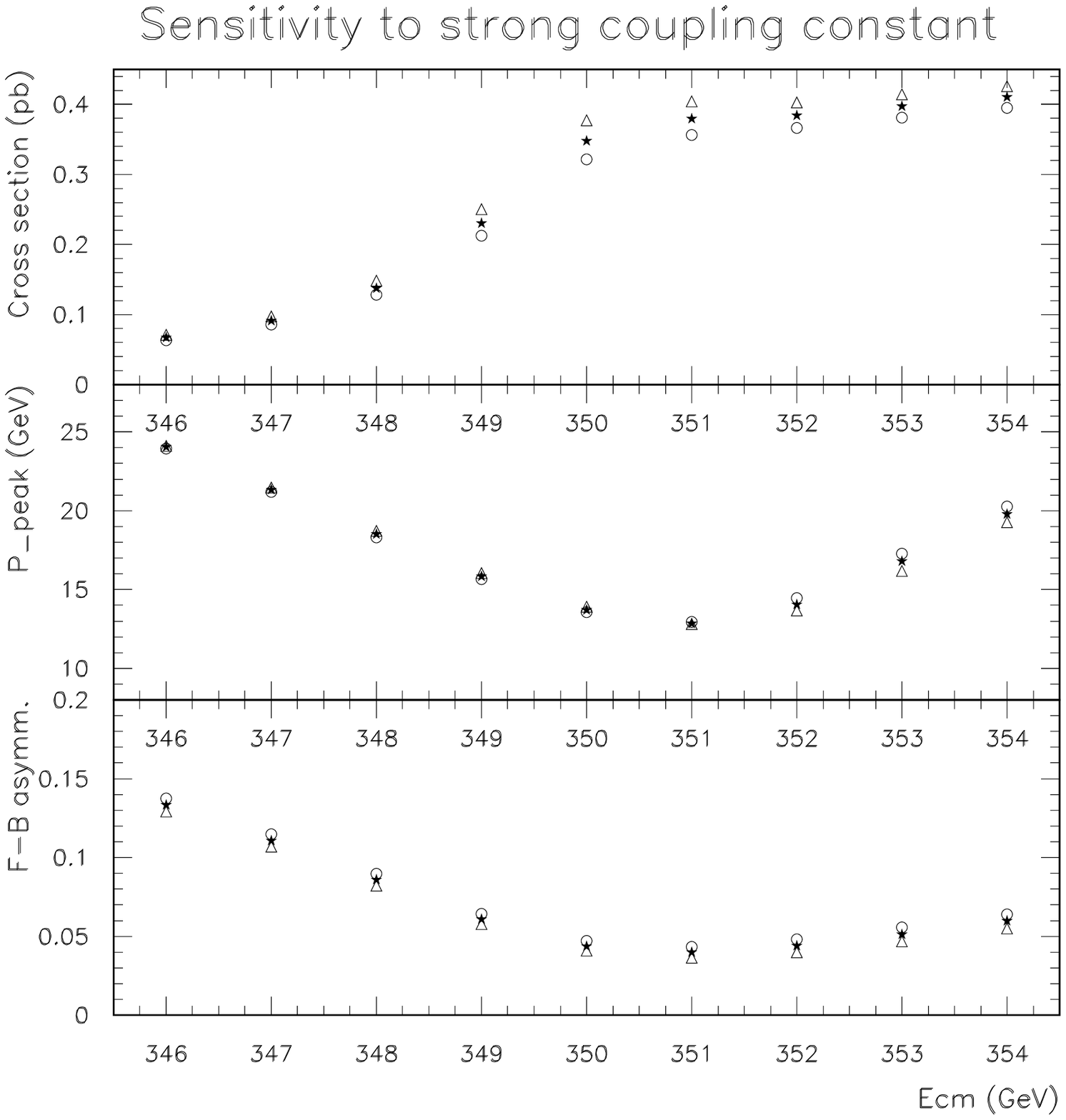}
\caption{Sensitivity of the observables to the strong coupling constant. 
The different markers correspond to $\Delta \alpha_s = 0.004$ intervals.}
\label{fig:sensitivity-as}
\end{figure}

\begin{figure}
\epsfxsize=15cm
\leavevmode
\centering
\epsffile{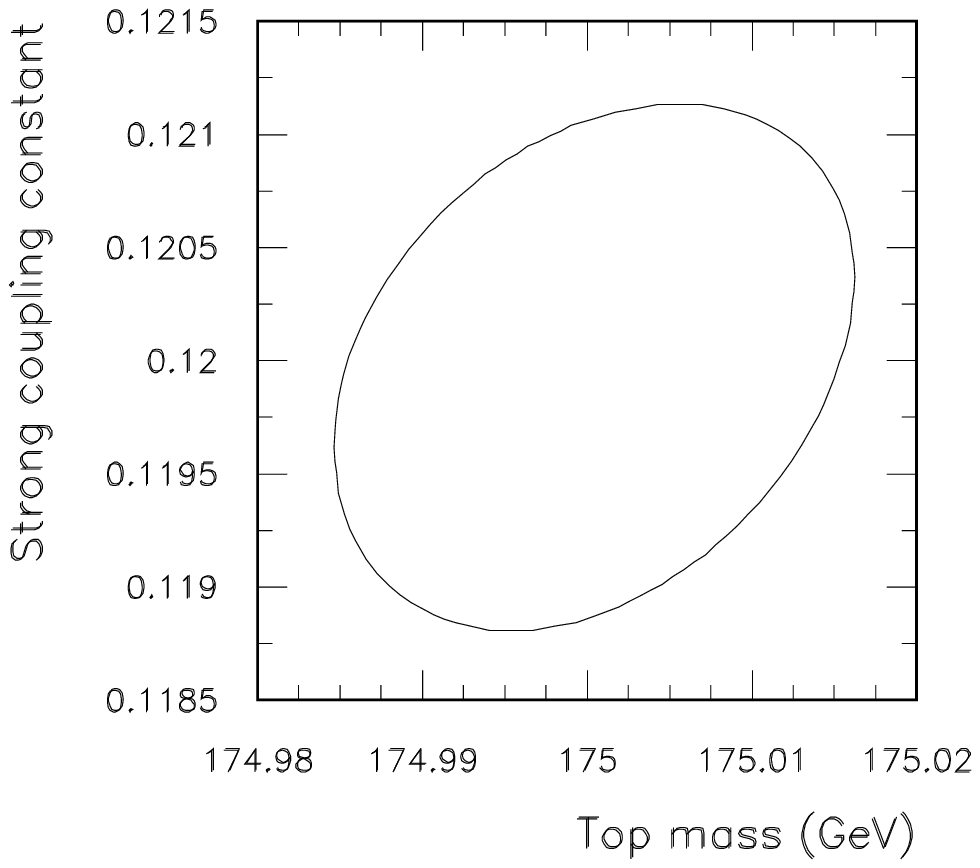}
\caption{$\Delta\chi^2=1$ contour as a function of $m_t(1S)$ and $\alpha_s(M_Z)$.}
\label{fig:mtVsAs}
\end{figure}

The sensitivity of each one of the observables to the top mass can be 
gleaned from fig.~\ref{fig:sensitivity-mt}
while the sensitivity to the strong coupling constant can be seen in fig.~\ref{fig:sensitivity-as}. 
Given the scale of the experimental uncertainties shown in the previous section, 
it is easy to see that no relevant information 
in these two parameters can be expected from the forward-backward asymmetry. 
The peak of the momentum distribution is fairly sensitive to 
the top mass while rather insensitive to $\alpha_s$ and, even within the 
modest experimental precision, should provide valuable 
information on $m_t$. Finally, the cross section is very sensitive to both the 
top mass and $\alpha_s$ but, since we are using the 
$1S$ top mass definition, while most of the sensitivity to the top mass is in the 
(smeared) threshold position, the sensitivity 
to $\alpha_s$ comes mainly from the cross section above threshold. This indicates already 
that the top mass determination would benefit from 
investing the luminosity of the scan points which are above threshold in the 
threshold rising slope instead. 
The resulting uncertainties, including only experimental errors, are the following:
\begin{equation}
\Delta m_t = 16\ {\rm MeV} \ \ \ \ \ \ \Delta\alpha_s = 0.0011 \ \ \ \ \ \  \rho = 0.36 \ ,
\end{equation}
where $\rho$ is the correlation coefficient between $m_t$ and $\alpha_s$. 
If the 3\% theoretical normalization error in the cross section is included, the results change to:
\begin{equation}
\Delta m_t = 16\ {\rm MeV} \ \ \ \ \ \ \Delta\alpha_s = 0.0012 \ \ \ \ \ \  \rho = 0.33 \ ,
\end{equation}
As it can be seen, the change is extremely small. In the rest of the paper, unless explicitly said otherwise, 
all numbers and figures given will include the effect of the theoretical error.
Figure~\ref{fig:mtVsAs} shows the correlation plot between the top mass and $\alpha_s$ resulting
from the two-parameter $\chi^2$ fit.

To quantify the actual contribution of the different observables to this two-parameter fit, it 
has been repeated using only the cross section. In this case the results are
\begin{equation}
\Delta m_t = 25\ {\rm MeV} \ \ \ \ \ \ \Delta\alpha_s = 0.0019 \ \ \ \ \ \  \rho = 0.76 \ ,
\end{equation}
The difference between this result and the one above has been traced back to
the contribution of the peak of the momentum distribution, as expected. 
The sensitivity of the momentum distribution is such
that it gives raise to a negative correlation between the extracted values of $m_t$ and
$\alpha_s$ which, when combined with the positive correlation coming from the cross
section measurement, leads to substantial reductions in the overall errors of both
$m_t$ and $\alpha_s$.

Once the possibility of measuring precisely the top mass and $\alpha_s$ with a $\bar{t}t$ 
threshold scan has been established one may turn the attention to measuring other quantities,
like the top quark width and the top quark Yukawa coupling.
\section{Multi-parameter fit strategy}
So far, only single-parameter (or at most two-parameter) fits have been tried to the threshold 
scan observables to study the potential for the determination of the top width and Yukawa coupling.
Nevertheless, as we have seen, the expected experimental error in each one of the observables strongly 
suggests that very likely
the cross section measurement will dominate the parameter determination (with only a small improvement 
eventually expectable 
from adding the information from the forward-backward asymmetry and the top momentum distributions). 

Therefore, as already stressed above, non-negligible correlations might be expected between the four parameters 
(top mass, strong coupling constant, top width and top Yukawa coupling) and, in this scenario, the only valid 
approach is a multi-parameter fit strategy.

For that, we need predictions of the three threshold observables studied in this work as a function of the four free
parameters. In spite of the 
fact that the TOPPIK code speed has sizably improved over the previous versions, the time needed to run it after 
the convolution of the predictions with ISR and with the beam spectrum makes its use for a 4P fit impossible. 
To cure this problem and produce the necessary predictions in an affordable amount of time we have used a  
multidimensional interpolating routine based upon the algorithms of ref.~\cite{Numerical-Recipes}.
We have checked that, within the parameter intervals relevant for the 
fits discussed in this work, that interpolation produces results which reproduce the exact predictions
with the required accuracy.

\section{The Top Quark Width}
Earlier attempts have been made in studying the determination of the top quark width ($\Gamma_t$) 
from the $\bar{t}t$ 
threshold scan~\cite{Comas, afb-sumino, gt-old-jp, gt-old-us}, 
with results in less than perfect agreement with each other, at least apparently.

\begin{figure}
\epsfxsize=15cm
\leavevmode
\centering
\epsffile{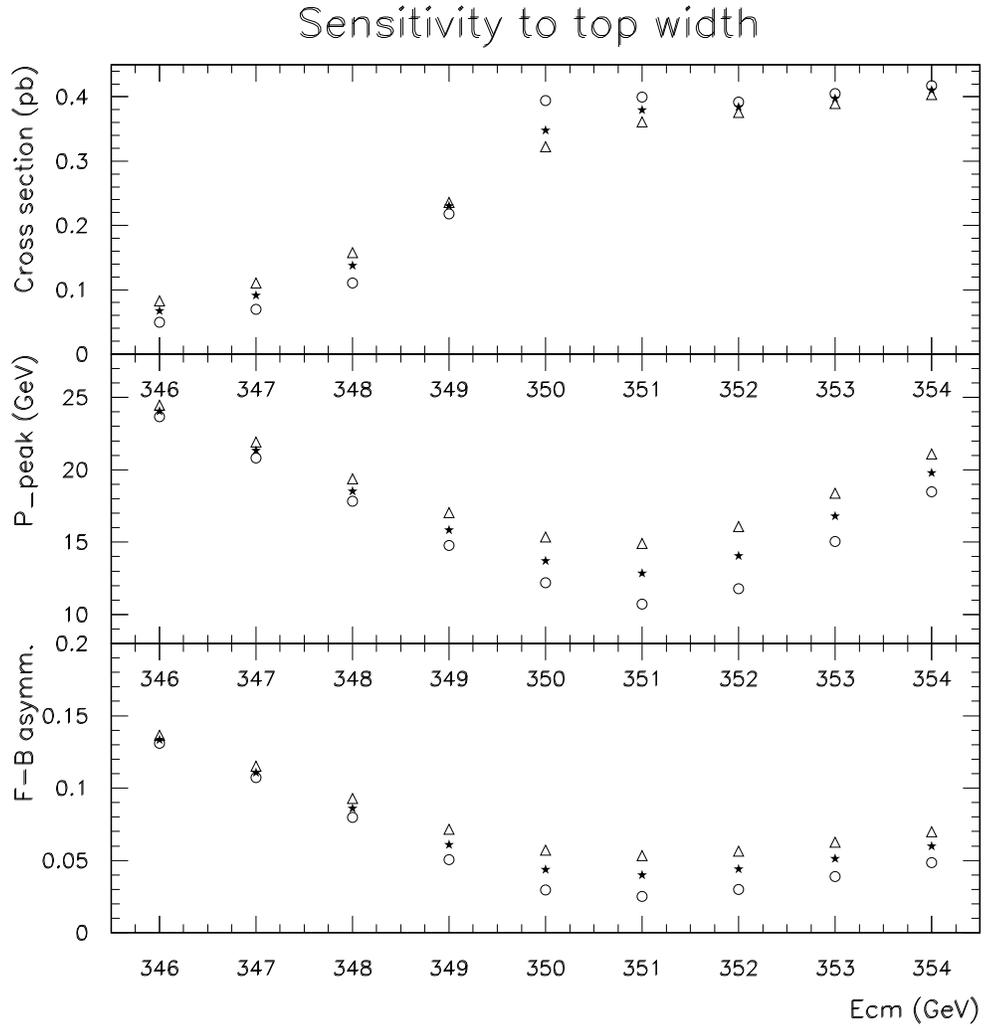}
\caption{Sensitivity of the observables to the top width. 
The different markers correspond to $\Delta \Gamma_t = 400$ MeV 
intervals.}
\label{fig:sensitivity-gt}
\end{figure}

Figure~\ref{fig:sensitivity-gt} shows the sensitivity of the threshold observables to the top width.
As it can be seen from the figure, there is sizable sensitivity both at the peak of the 
cross section and at lower center-of-mass energies. For low values of the top width, the peak structure 
of the $1S$ resonance becomes apparent while for large values it disappears. The other two observables ($A_{FB}$ and
peak of momentum distribution) have an even larger sensitivity, which is enhanced in the energy points above threshold. 
Unfortunately, as already stressed, the accuracy of their experimental determination is much poorer. Because 
of that, in the scenario presented in this work, the determination of the top width can be expected to be 
dominated by the cross section measurement. 

\begin{figure}
\epsfxsize=15cm
\leavevmode
\centering
\epsffile{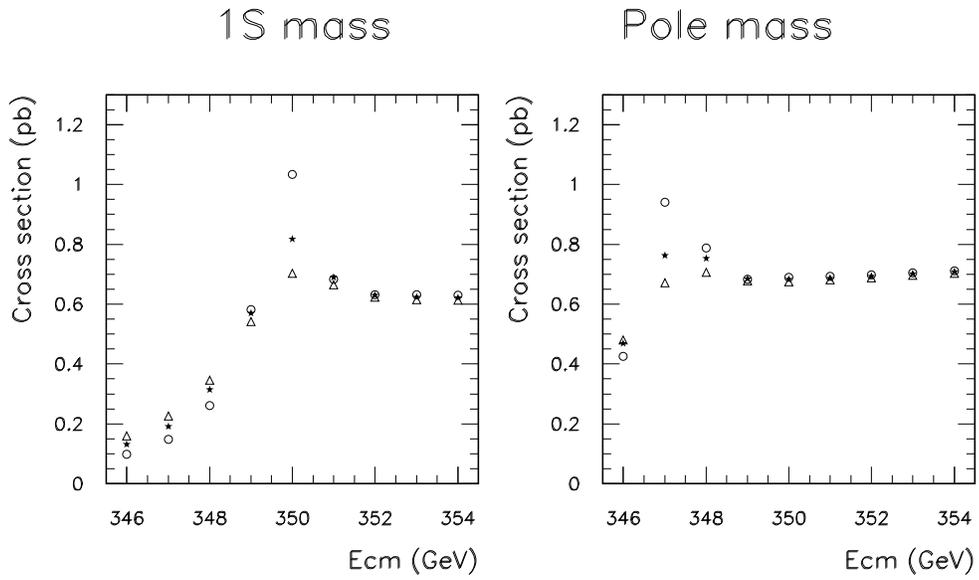}
\caption{The sensitivity of the cross section prediction (without ISR or beam effects) to the top 
width variation when the scan is centered around the top 1S-mass and when it is centered around 
the top pole-mass. The different markers correspond to $\Delta \Gamma_t = 400$ MeV 
intervals.}
\label{fig:width-comparison}
\end{figure}

A three-parameter fit,
with $m_t$, $\alpha_s$ and $\Gamma_t$ leads to the following overall uncertainties:
\begin{equation}
\Delta m_t = 19\ {\rm MeV} \ \ \ \ \ \ \Delta\alpha_s = 0.0012 \ \ \ \ \ \  
\Delta\Gamma_t = 32\ {\rm MeV} \ ,
\end{equation}
with all correlations between the three parameters being below 50\%. The $32$ MeV uncertainty on
the top width corresponds to about a 2\% measurement. This has to be compared with the 18\%
uncertainty reported on ref.~\cite{Comas}. Several factors account for the decrease in the 
reported error:
\begin{itemize}
\item The integrated luminosity assumed now is six times larger than that assumed 
in~\cite{Comas} ($30$ $fb^{-1}$ instead of $5$ $fb^{-1}$ per point). This accounts for a 
factor $\sim 2.4$.
\item The selection efficiency for top-antitop events has been taken from the work 
in~\cite{Aurelio}, and it is about 41\% while the one used in~\cite{Comas} was of about 25\%. 
This accounts for a factor $\sim 1.3$.
\item The present studies are based on a newer version of the TESLA machine. 
As a consequence the beam spectrum is now sharper than previously assumed. This results in an 
improvement of the top width determination by a factor $\sim 1.5$.
\item A substantial part of the sensitivity in the center-of-mass energies below the maximum
of the cross section was lost 
in the scan of~\cite{Comas}, because, although the scan was centered around $2m_t$, the
position of the peak in the cross section was shifted by about 2~GeV toward lower $\sqrt s$,
because of the pole mass definition used for $m_t$. In contrast, both PS and 1S masses result,
essentially by definition, in a cross section with a maximum at $2m_t$, so that the scanning
strategy used here catches those energies below the maximum with substantial sensitivity to
the top width. This difference can be clearly seen in figure \ref{fig:width-comparison}. 
This accounts for a factor $\sim 1.8$.
\end{itemize}

When all these changes are taken into account together, a factor $\sim 8.5$ difference is 
obtained and therefore 
good compatibility is found between the results presented here and those in~\cite{Comas}. 

To disentangle the contribution of the different observables to the top width determination, the fit has been
repeated using only the cross section. The results are
\begin{equation}
\Delta m_t = 34\ {\rm MeV} \ \ \ \ \ \ \Delta\alpha_s = 0.0023 \ \ \ \ \ \  
\Delta\Gamma_t = 42\ {\rm MeV} \ ,
\end{equation}
with correlations as large as 80\% between the top mass and $\alpha_s$.
The difference between this fit and the result above can be traced back completely to the 
contribution of the peak of the momentum distribution in determining $m_t$. The contribution of 
the forward-backward asymmetry, introduced because conceptually should be the cleanest observable to
see the effect of the top width due to the overlap between the $1S$ and $1P$ states is in practice 
negligible.

It is important to stress here that, being the top quark so heavy,
a 2\% determination of the top quark width
can be very useful in constraining models of new physics which would predict new particles that
could be produced on top quark decays. The precise determination of the top width from the
threshold scan allows to put constraints which are independent of the final state particles
produced.
\section{The Top Yukawa Coupling}
\begin{figure}
\epsfxsize=15cm
\leavevmode
\centering
\epsffile{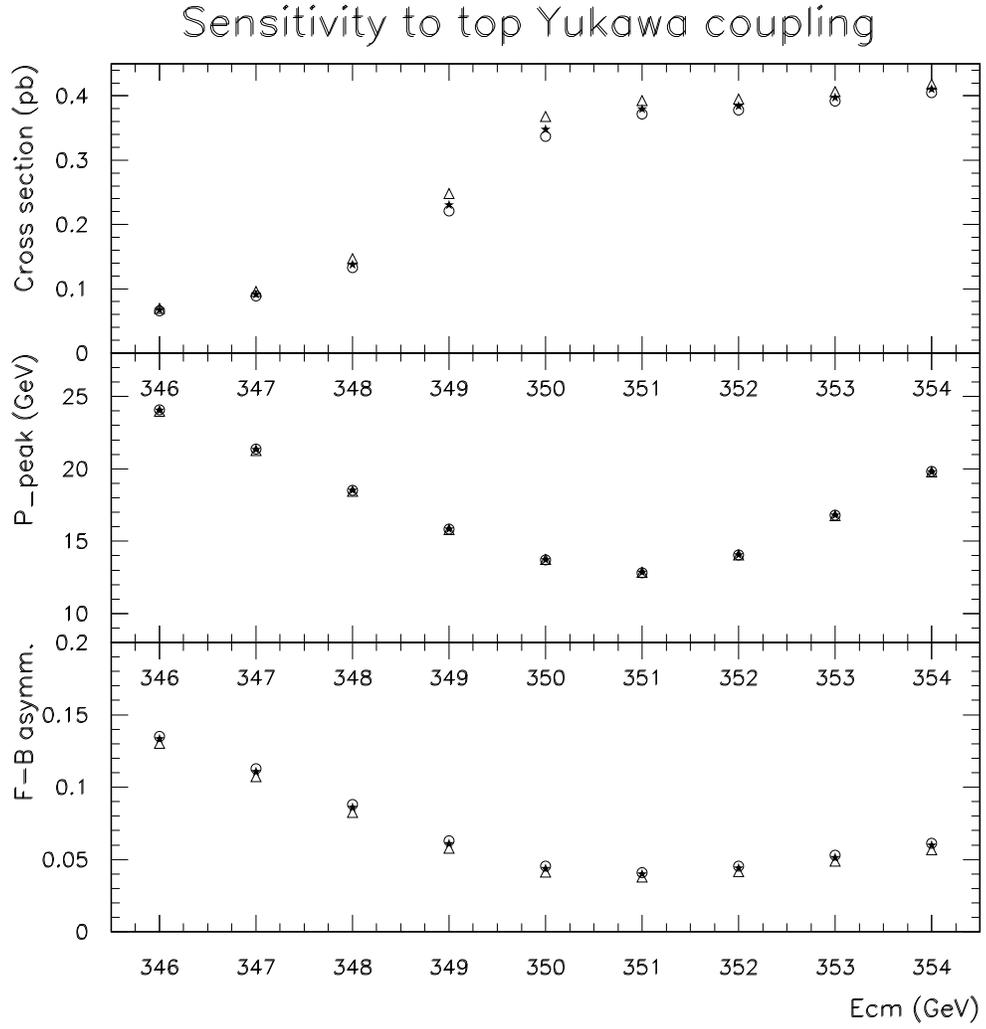}
\caption{Sensitivity of the observables to the top Yukawa coupling. The different markers correspond 
to $\Delta \lambda_t / \lambda_t = 0.50$ intervals.}
\label{fig:sensitivity-yt}
\end{figure}

Measuring the top Yukawa coupling could provide an important test of the Higgs mechanism for
generating fermion masses. The exchange of a Higgs boson between the top and anti-top produced
at threshold has been taken into account in the theory prediction by adding a Yukawa potential to
the QCD $\bar{t}t$ potential~\cite{kuhn-yuk}. 
The modified potential can have measurable effects in the
observables studied here. However,
figure~\ref{fig:sensitivity-yt} shows that the sensitivity of the total cross
section to the Yukawa coupling is not very large and is not better in the forward-backward 
asymmetry while is unexisting in the peak of the momentum distribution. Following the same arguments 
given in the previous sections, we can expect that the Yukawa coupling determination will be completely
dominated by the cross section measurement as well.

As an unrealistic starting point, a
one-parameter fit is performed, fixing all parameters except for the top Yukawa coupling,
$\lambda_t$. The fit returns an asymmetric uncertainty:
\begin{equation}
\frac{\Delta\lambda_t}{\lambda_t} = ^{+0.18}_{-0.25} \ .
\end{equation}
Given the lack of sensitivity, one can try to see whether there could be any gain obtained by 
relaxing somewhat the assumptions concerning systematic errors. In particular, the systematic
error in the cross section determination has been lowered from 3\% (taken from~\cite{Aurelio})
to 1\%, which seems like a reasonable educated guess, given the level of understanding
achieved at electron-positron machines like LEP, where selection
systematics routinely achieved the
few per mil level. Also, maybe with less justification, the theoretical error in the overall
normalization has been lowered from 3\% to 1\%.
Assuming the 1\% errors, the uncertainty in the one-parameter fit decreases
to:
\begin{equation}
\frac{\Delta\lambda_t}{\lambda_t} = ^{+0.14}_{-0.20} \ .
\end{equation}
>From now on, these lower systematic errors in selection and normalization will be assumed on all fits.

The next step consists on leaving the top mass and $\alpha_s$ free in the fit while fixing the
top width to its Standard Model value and including an external constraint on $\alpha_s(M_Z)$
of $\pm 0.001$. The constraint should come from a different determination of $\alpha_s$, like
the one available, for instance, at GigaZ~\cite{moenig-as}.
Under these conditions, a three-parameter fit leads to the following precisions:
\begin{equation}
\Delta m_t = 27\ {\rm MeV} \ \ \ \ \ \ \Delta\alpha_s = 0.001 \ {\rm (constraint)} \ \ \ \ \ \ 
\frac{\Delta\lambda_t}{\lambda_t} = ^{+0.33}_{-0.54} \ ,
\end{equation}
with correlations large, up to 80\%, particularly among $m_t$ and $\lambda_t$. 

Finally, one could also try to leave the top width free in the fit, and perform a
four-parameter fit with an external constraint on $\alpha_s(M_Z)$. The results are:
\begin{eqnarray}
\Delta m_t &=& 31\ {\rm MeV} \ \ \ \ \ \ \ \ \Delta\alpha_s = 0.001 \ {\rm (constraint)} 
\nonumber \\ 
\Delta\Gamma_t &=& 34\ {\rm MeV} \ \ \ \ \ \ \ \ \frac{\Delta\lambda_t}{\lambda_t} = 
^{+0.35}_{-0.65} \ .
\end{eqnarray}
The simultaneous determination of the four parameters is possible without a large increase in the
resulting uncertainties. Correlations remain similar, with a maximum of 83\%, again among
$m_t$ and $\lambda_t$. 

As can be seen, a realistic determination of the top Yukawa coupling, which has to be done
simultaneously with that of the top mass, is very challenging, although not impossible.

\section{Summary}
For the first time a simultaneous 4-parameter (top mass, strong coupling constant, top width and top 
Yukawa coupling) fit to the expectations for three $t \bar{t}$ threshold scan observables (cross section, 
top momentum distribution and top forward-backward charge asymmetry) has been carried out.

In a complete 4-parameter fit to the expected threshold observables for a total 300~fb$^{-1}$ luminosity scan 
using the TESLA machine, the parameter correlations obtained are very important (up to 83\%), 
justifying the use of a full multidimensional approach.

The expected experimental uncertainties in each observable, as well as the 
sensitivity of each observable to the fitting parameters, have been scrutinized to understand in detail
the origin of the correlations obtained in the parameter determination.

The outcome of the fits shows that the prediction of a high precision determination of the top mass, with an 
experimental 
accuracy better that $30$ MeV and of the top width, with an accuracy at the 2\% level, is quite robust.

Measuring the top Yukawa coupling with a top threshold scan looks difficult. Even with somewhat optimistic 
assumptions, the best error which can be expected with the above luminosity is above 30\%, assuming a 
Higgs 
mass of $120$ GeV. The situation should become significantly worse for heavier Higgses.

For the first time, an estimate of the theoretical error in the cross section prediction (a 3\% normalization error
from~\cite{teubner2001}) has been included in the fits and has been shown to have little effect on the overall result.
However, a more realistic theoretical error model might lead to more severe errors, particularly on the top mass 
and $\alpha_s(M_Z)$.
%
\section*{Acknowledgments}
We would like to thank Thomas Teubner for making his calculations readily available to us and for many discussions
on the subject of this paper. The work presented here is indebted to previous work by Pere Comas, 
Aurelio Juste, Salvador Orteu and Dani Peralta. To them, our thanks. Over the period of over ten years
in which we have been working intermittently
on this topic we have benefited from discussions with many people. We would like to thank them all and
mention in particular Peter Igo-Kemenes, Hans K\"uhn, Marek Je\.{z}abek, David Miller and Daniel Schulte.
Finally, we want to thank also Andre Hoang, Antonio Pineda, Joan Soto and Iain Stewart for providing us with detailed 
information about
the latest theoretical developments on the subject.

R.M.~is supported by the U.S.~Department of Energy under contract no.~DE-AC03-76SF00098 and by the U.S.~National Science
Foundation under agreement no.~PHY-0070972.

\end{document}